\documentclass[12pt]{iopart}

\usepackage{dcolumn}
\usepackage{amssymb}
\usepackage{graphicx}
\usepackage{epsf,epsfig}
\usepackage{epstopdf}
\usepackage{color}

\usepackage{color}

\newcommand \be {\begin{equation}}
\newcommand \ee {\end{equation}}
\newcommand \bea {\begin{eqnarray}}
\newcommand \eea {\end{eqnarray}}

\begin{document}

\title{Dynamical fluctuations in a simple housing market model}

\author{R\'emi Lemoy and Eric Bertin}

\address{Universit\'e de Lyon, Laboratoire de Physique, ENS de Lyon, CNRS,
46 All\'ee d'Italie, 69007 Lyon, France}

\eads{\mailto{remilemoy@gmail.com}, \mailto{eric.bertin@ens-lyon.fr}}

\date{\today}

\begin{abstract}
We consider a simple stochastic model of a urban rental housing market, in which the interaction of tenants and landlords induces rent fluctuations.
We simulate the model numerically and measure the equilibrium rent distribution,
which is found to be close to a lognormal law.
We also study the influence of the density of agents (or equivalently, the vacancy rate)
on the rent distribution. 
A simplified version of the model, amenable to analytical treatment, is studied
and leads to a lognormal distribution of rents.
The predicted equilibrium value agrees quantitatively with numerical simulations,
while a qualitative agreement is obtained for the standard deviation.
The connection with non-equilibrium statistical physics models like ratchets is also
emphasized.
\end{abstract}


%\pacs{89.75.Fb}{Structures and organization in complex systems}
%\pacs{02.50.Ey}{Stochastic processes}
%\pacs{89.65.Lm}{Urban planning and construction}

%\maketitle

\section{Introduction}

The emergence of a well-defined distribution of prices in a market involving a large number
of interacting agents is a typical example of collective phenomenon in complex systems,
for which concepts and tools borrowed from non-equilibrium statistical physics could prove
useful.
Theoretical studies of price fluctuations have been reported both in the economic \cite{RamaBouchaud2000} and in the physics literature, mostly in the framework of stock market models, where price fluctuations are large,
with often fat-tailed distributions \cite{Bak97,Caldarelli97,Sato98}.
Other markets, like the rental housing market or the labor market, seem to have attracted less attention from the physics community
(see, however, \cite{Watanabe2010,Nadal2010}), though their potential interest from a modeling point of view is certainly high.
The economic literature on housing market is indeed in close relationship with the literature on labor market, which inspired some of the first models in the domain \cite{Arnott1989, Wheaton1990}. These works study processes of search and matching between tenants (or persons wishing to purchase a house on the residential property market) and housing lots and their owners. The importance of the vacancy rate for the turnover between tenants and its influence on price formation have been underlined. Several works have studied price dispersion, and the conditions needed to model it on the housing market \cite{McMillan1988, Read1991, Yavas2001}, or corresponding wage dispersion on the labor market \cite{Burdett1998, Hornstein2011} and general price dispersion for consumer goods \cite{Burdett1983}. The phenomenon of price dispersion is also documented in the empirical literature, be it for housing market \cite{Leung2006,McMillen08} or prices of products in stores for instance \cite{Lach2002}.

The mechanisms at the origin of price dispersion in economic models are often different
from that responsible for fluctuations in physical systems, namely thermal fluctuations.
In most economic models, there is an exogenous heterogeneity in agents' behavior rules \cite{Burdett1983}:
firms producing goods can have different costs of production, goods can have different characteristics \cite{Read1991},
or consumers may have different search costs or behaviors \cite{Yavas2001}.
The literature on search and matching theory focuses on the formation of relationships \cite{Burdett1983, Wheaton1990, Burdett1998, Pissarides2000}, between buyers and sellers for instance. There price dispersion usually comes from a difference in the information agents have on the market, or a difference in access to opportunities
(e.g., a potential tenant is presented only a few flat offers among many).

Departing from this standard economic framework, we propose a parcimonious model of urban rental housing market, where the behavior of landlords and tenants is probabilistic, and driven by very limited information.
{\color{black} Indeed, as explained in detail below, agents have in this model a very limited knowledge of the market, at odds with the "optimizing agent" framework often used in the search and matching literature. For instance, tenants only know their current rent, and the rent in a randomly chosen flat where they consider moving. Another important difference with the "optimizing agent" framework is our use of probabilistic rules for moves.}
We study both numerically and analytically this simple stochastic model, in order to characterize the rent distribution, and to determine the role of the density of agents (considered to be an external control parameter) on the rent formation mechanism. 
From a statistical physics viewpoint, our model also presents some similarities with non-equilibrium systems like ratchet models driven by dichotomous Markov noise \cite{bena06}, suggesting interesting connections between the two fields.

\section{A simple model of rental housing market}
\label{EvoPrix_modele}

The model simulates the evolution of the distribution of rents in a city. {\color{black} Note that we focus on a rental housing market, and do not treat the question of the relation between rents and housing prices.}
In the model, the virtual city is composed of $N_l$ flats labeled by $i=1,...,N_l$, which are supposed to have identical characteristics, except for their rent, and $N_a < N_l$ agents, with $N_a/N_l=\rho$ the density of agents. The flats can be in only two states: either occupied by an agent, or empty. When a flat is occupied, there is a probability for the tenant residing in it to move out, and when a flat is empty, there is a probability for a tenant to move in. 

In addition, the rent increases when a flat is occupied, as the rent is revised periodically by the owner. When a flat is empty, its rent decreases until some agent moves in. These ingredients give a simple dynamical framework to ensure a competition on the housing market as in standard urban economics models \cite{Fujita89}, and are consistent with the fact that demand and supply drive the rent.
From a physics point of view, this model is analogous to a lattice gas with long-range hopping, but is however more complicated due to the coupling with a dynamically evolving rent. 

The model is simulated in discrete time. At each time step, a flat is chosen at random in the city; this flat is either occupied or empty. The other variable characterizing the flat is its rent $p$, i.e., the rent that a tenant pays or would need to pay to occupy the flat. Let us first consider the case where the chosen flat is occupied.
In this case, there is a given probability $\pi_r$ that the rent is raised by the owner during the considered time step. If the rent is increased, it is multiplied by a given factor $f_r>1$. 
In a second substep, there is a probability $\pi_s$ that the tenant searches for another flat in the city with a lower rent. In this case, an empty flat, whose rent is denoted $p_e$, is chosen at random in the city and the agent moves into this flat only if $p_e<p$, the initial flat thus becoming empty.
If $p_e>p$, the agent does not move, and the selected empty flat is rejected.
The probability $\pi_s$ depends linearly on the rent $p$ of the considered flat: $\pi_s = \min(p/p_s,1)$, where $p_s$ is a characteristic rent, which can be linked to the reservation rent in standard housing economics works. When the rent of the flat $p$ is high, that is to say, close to or above $p_s$, the probability that the tenant looks for another flat is high.

\begin{figure}[t]
\begin{center}
\includegraphics[width=9cm]{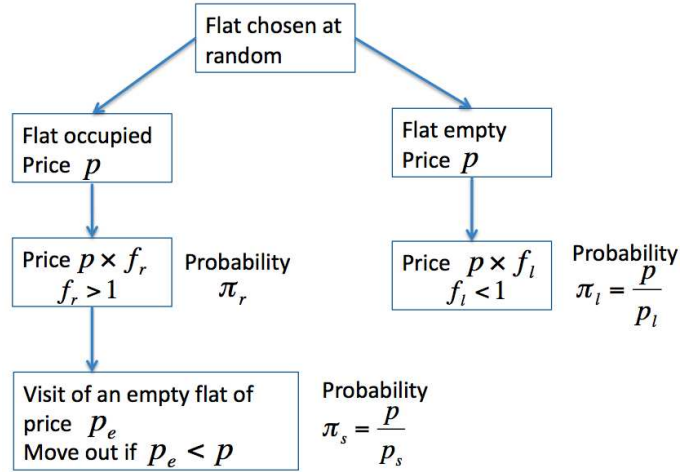}
\caption[Summary of the dynamics of the model.]{Summary of a time step of the model. Note that $\pi_s=1$ if $p/p_s > 1$, and $\pi_l=1$ if $p/p_l > 1$ (see text).}
\label{day}
\end{center}
\end{figure}

On the other hand, if the flat chosen at the beginning of the time step is empty, the rent is lowered by the owner with a probability $\pi_l$, in order to make the flat more attractive for potential tenants.
At the end of the time step, the flat remains empty.
This probability also depends on the rent $p$ of the flat: $\pi_l = \min(p / p_l,1)$, where $p_l$ is another characteristic rent, which could be related to rental provision costs in housing economics \cite{Read1991}. In case the rent is lowered, it is multiplied by a given factor $f_l<1$. {\color{black}Let us note that the behavior of landlords is also parcimonious. In order to optimize their profit, with a knowledge only of the state and rent of their flat, they raise the rent with a fixed probability when the flat is occupied (which can be interpreted as risk neutral behavior), but they are reluctant to decrease the rent when it is already too low.}
A summary of the dynamics of the model is given on Fig.~\ref{day}.

\section{Numerical simulations and results}

For practical reasons, and in order to facilitate comparison with the
analytical model studied in Sec.~\ref{analytical},
it is convenient to discretize the decimal logarithm of the rent $p$.
This discretization allows in particular for an easier determination
of the rent distribution.
We thus introduce a resolution $a \ll 1$, such that $\log_{10} p = na$,
or equivalently $p=10^{na}$, where $n$ is a positive integer.
For consistency, we also choose $f_r$ and $f_l$ in the form
$f_r=10^{n_r a}$ and $f_l=10^{-n_l a}$, with $n_r$ and $n_l$ two given positive integers.

\begin{figure}[t]
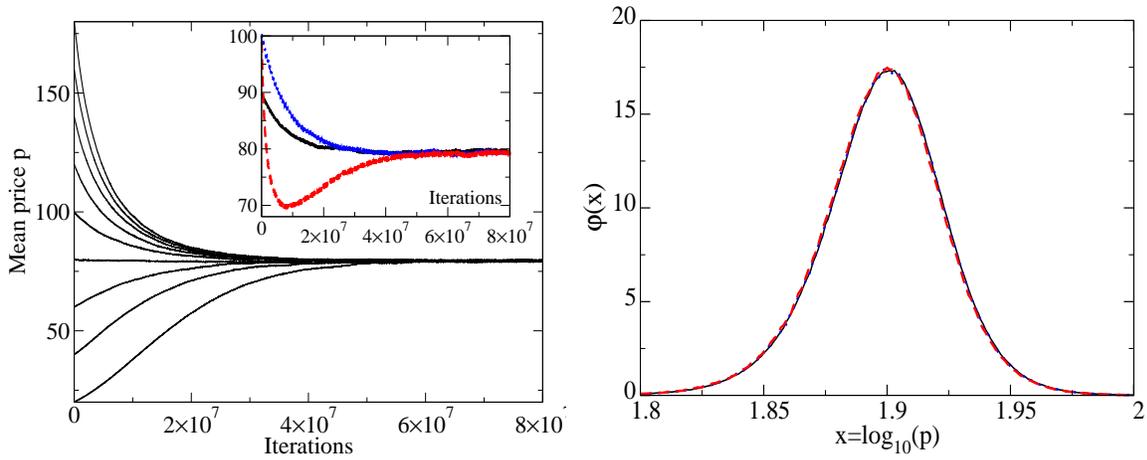

\begin{center}
\includegraphics[width=7.5cm, height=5.8cm]{EvolPrix1_v1.eps}
\includegraphics[width=7.5cm, height=6cm]{EvolPrix2_v1.eps}
\caption[Evolution of the mean rent over time]{Left panel: Evolution of the mean rent over time, with different initial rent distributions.
Main figure: Gaussian initial distribution of the rent, with standard deviation $5$ and mean $20$, $40$, $\ldots$, $180$.
Inset: evolution of the mean rent for initial distributions with different shapes.
Full line: Gaussian distribution of mean 90 and standard deviation 5;
dashed line: uniform distribution between 0 and 200;
dotted line: Dirac peak at p=100.
Parameters: $N_l=5000$, $p_s=p_l=2000$, $\pi_r=0.03$, $f_r=10^{11a} \simeq 1.026$, $f_l=10^{-19a} \simeq 0.957$ with $a=0.001$, $\rho=0.7$.
Right panel: equilibrium distributions of the decimal logarithm $x$ of rent for the same initial distribution
as on the inset of the left panel. All three distributions are identical
to numerical accuracy.}
\label{EvoPmean}
\end{center}
\end{figure}

The simulation is initialized as follows. Agents initially occupy the first $N_a$ flats $i=1,...,N_a$, and the $N_l-N_a$ remaining flats are empty. The rent of each flat is drawn at random from a normal distribution of given mean and variance. We find that the system reaches after some time an equilibrium rent that is independent of the initial rent distribution,
as illustrated on the left panel of Fig.~\ref{EvoPmean}.

To go beyond the average rent, and to characterize rent fluctuations, we measure
the equilibrium rent distribution, which is obtained by averaging the instantaneous rent histogram over a long time lag, once equilibrium is reached.
The resulting distribution is presented in the right panel of
Fig.~\ref{EvoPmean}, showing that the equilibrium distribution, and not only
the mean rent, is the same independently of the initial distribution.
Note that we actually measured the histogram of the logarithm of the rent,
to better visualize the relative dispersion of rents.

\begin{figure}[t]
\begin{center}
\includegraphics[width=7.5cm]{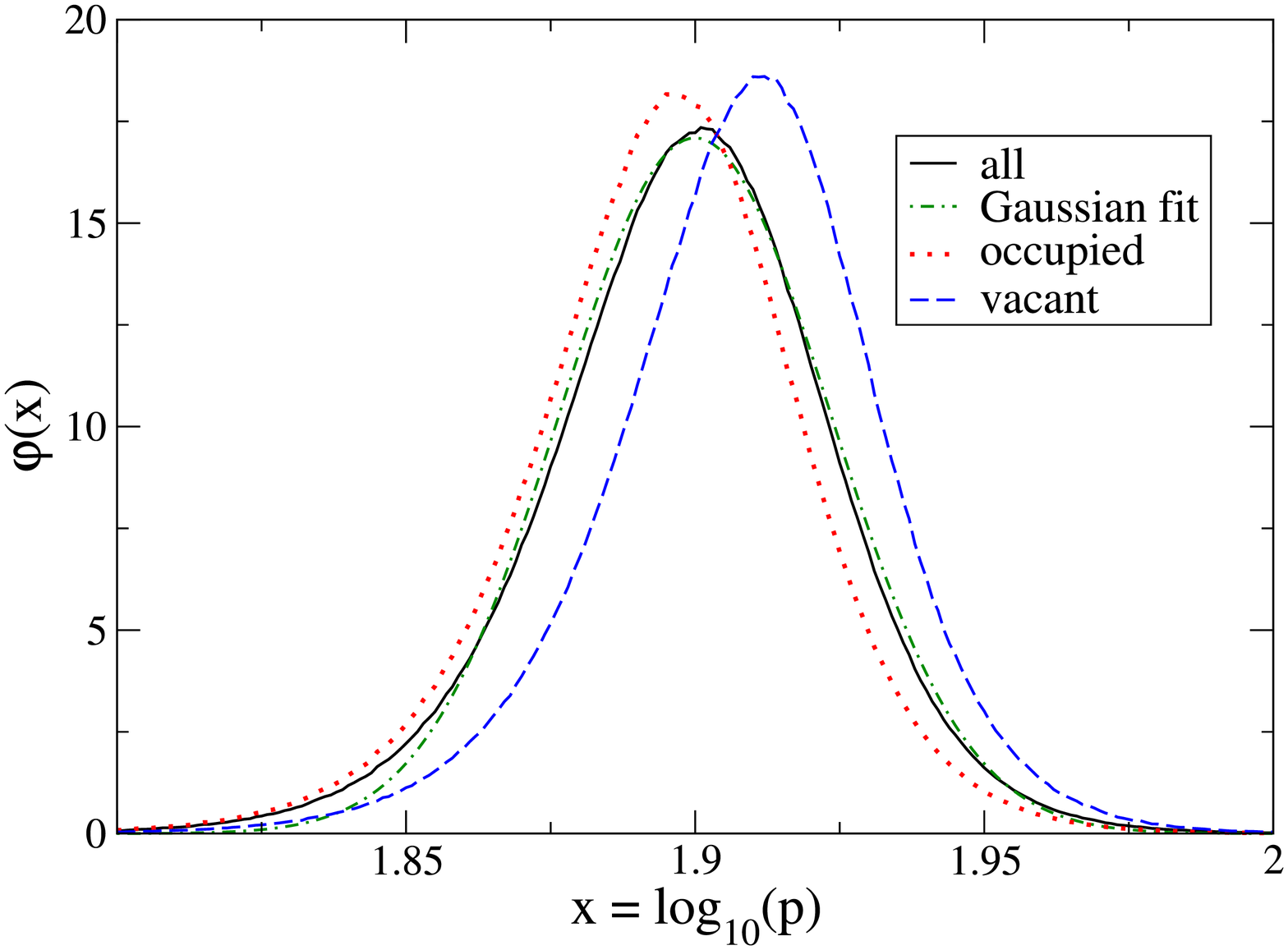}
\includegraphics[width=7.5cm]{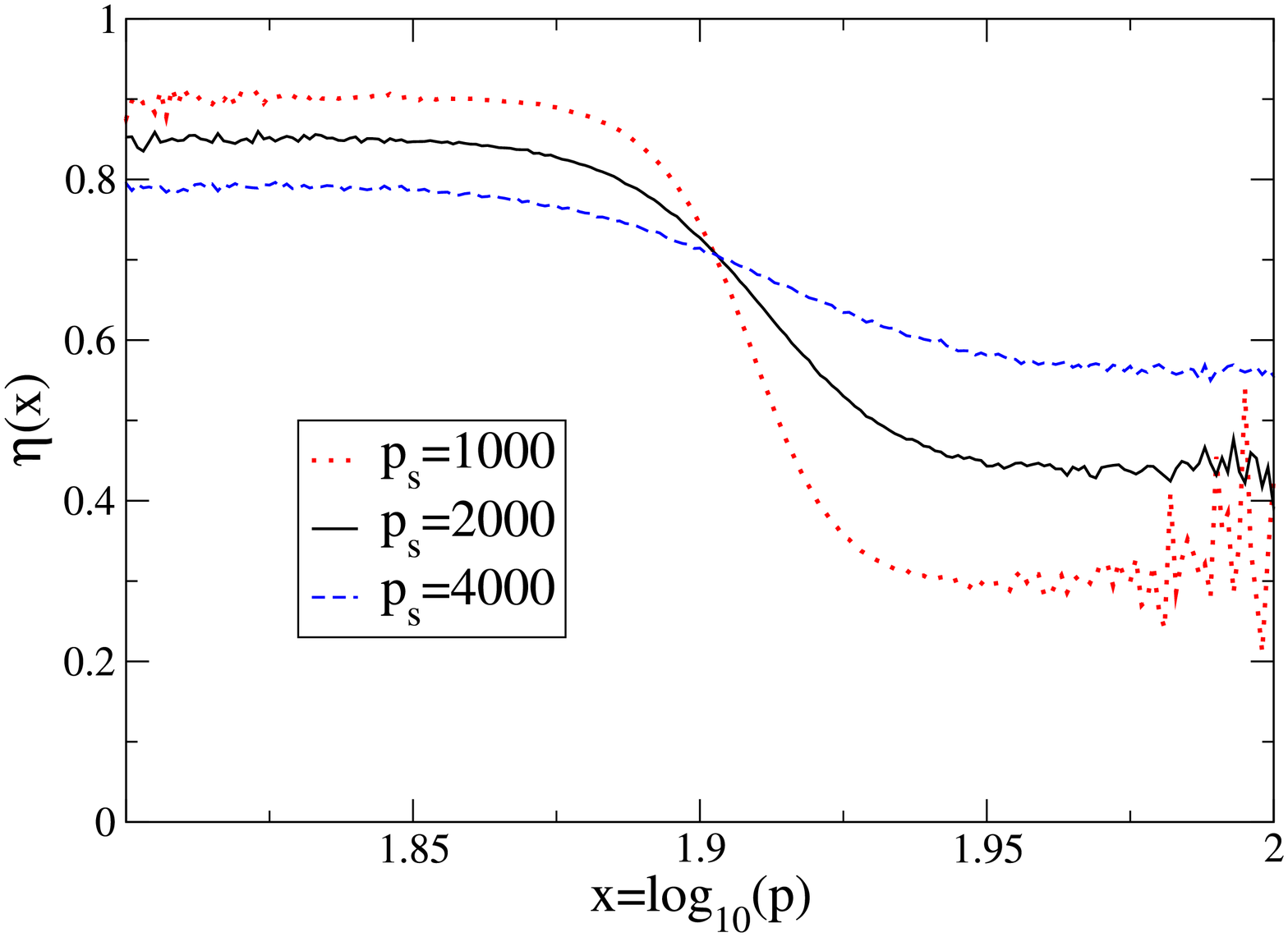}
\caption[Rent distribution and occupation rate of flats]{Left panel: averaged distribution $\varphi(x)$ of the logarithm $x$ of rent (for all flats, continuous line, for occupied flats only, dotted line, for vacant flats, dashed line) at equilibrium. Right panel: corresponding occupation rate $\eta(x)$ as a function of the logarithm $x$ of rent (continuous line). Same parameters as Fig.~\ref{EvoPmean}, except for the right panel where $p_s=1000$ for the dotted line and $p_s=4000$ for the dashed line.}
\label{prix_occup}
\end{center}
\end{figure}

The left panel of Fig.~\ref{prix_occup} presents separately the distributions of the logarithm of rent for occupied and vacant flats, which shows that demand, corresponding to occupied flats, is centered around inferior rents than supply (vacant flats), but both distributions intersect on a large range of rents, and build together the equilibrium rent distribution.
This interplay of demand and supply leading to equilibrium rent provides a qualitative link between our work and standard economic theory on rent formation \cite{MicroeconomicTheory}.

%It is also interesting to look at the distribution of (the decimal logarithm)
%of the rents of occupied and empty flats respectively.
%These distributions are shown in the top panel of Fig.~\ref{prix_occup},
%and compared with the global distribution of rents.

This feature of the model is also illustrated by another quantity of interest: the mean share of time during which a flat is occupied,
which characterizes the occupation of flats at equilibrium
(see right panel of Fig.~\ref{prix_occup}). This occupation rate illustrates the dynamics of the model: when rent is low, occupation rate is high, and vice versa. Occupation -- corresponding to demand -- is governing the rent of flats in this model where rent is the only variable which distinguishes housing lots in the city.

\begin{figure}[t]
\begin{center}
\includegraphics[width=7.5cm]{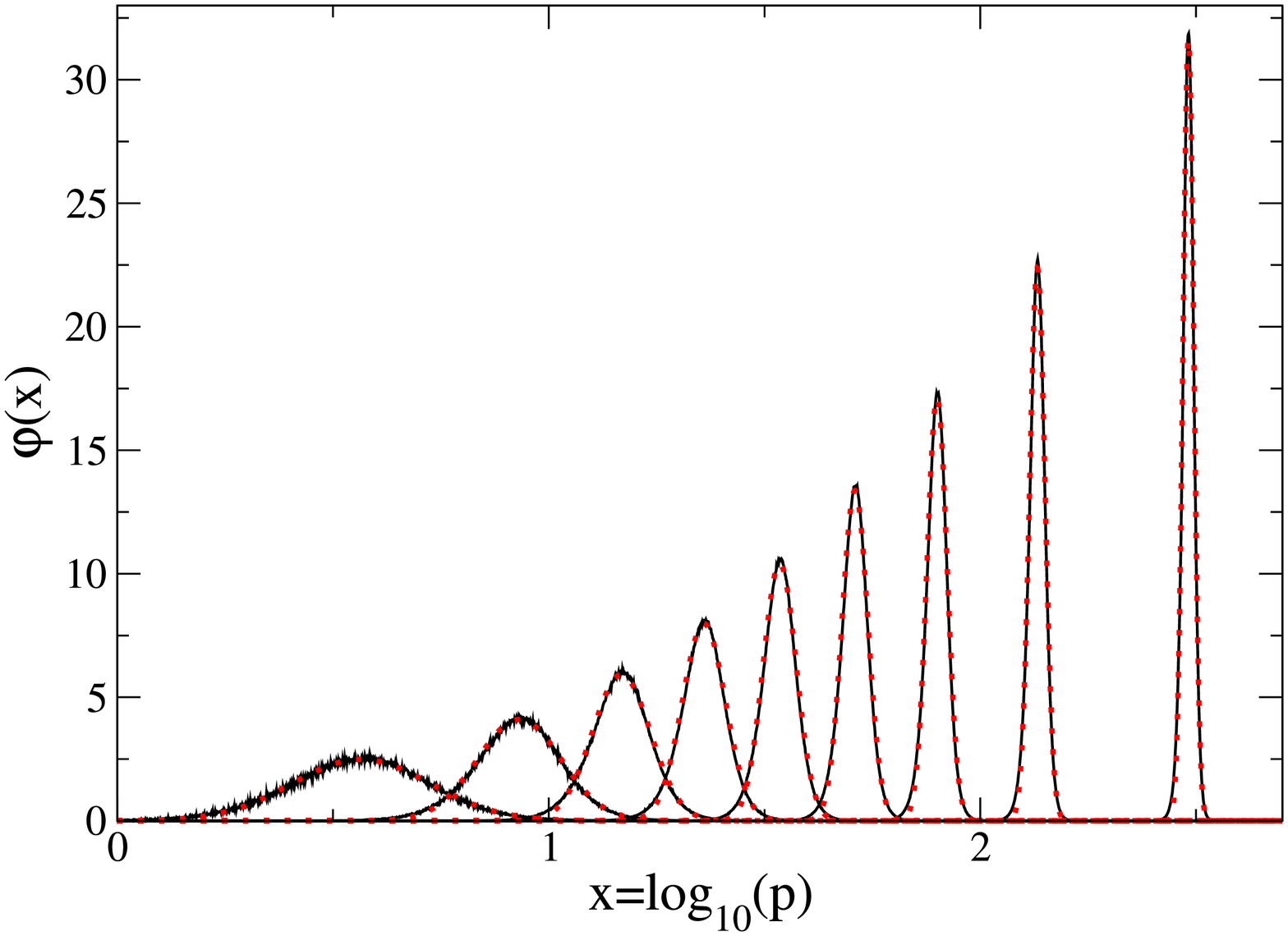}
\includegraphics[width=7.5cm]{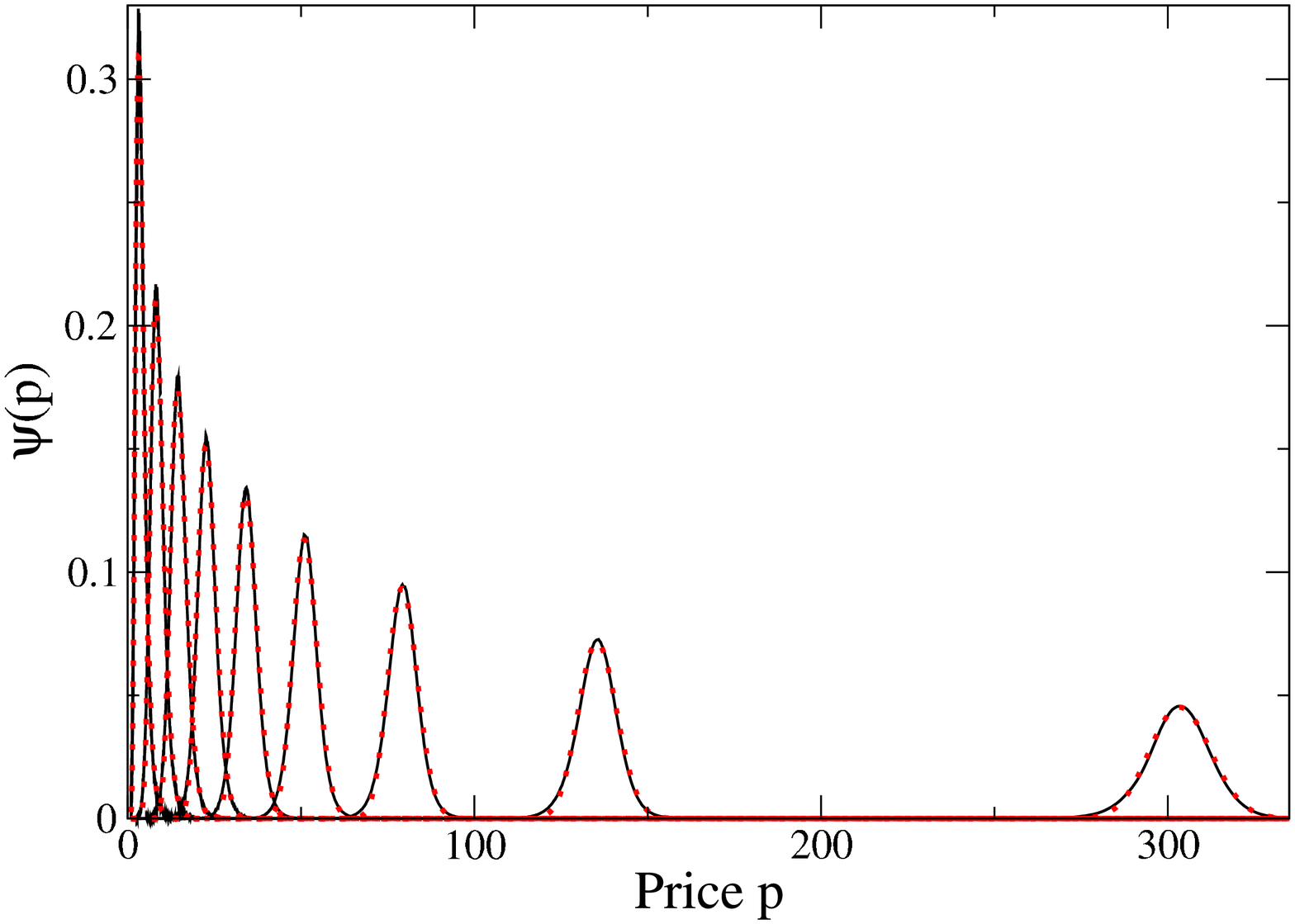}
\caption[Equilibrium distribution $\phi (x)$ of the decimal logarithm of rent, for different values of the global density $\rho$.]{Equilibrium distributions $\varphi(x)$ of the decimal logarithm of rent (left panel, continuous lines -- dotted lines are Gaussian fits) and $\psi(p)$ of rent (right panel, continuous lines -- dotted lines are lognormal fits) for different values of the global density $\rho$. Same parameters as Fig.~\ref{EvoPmean}, except for the density: $\rho= 0.1,0.2,...,0.9$ from left to right.}
\label{Prix_rho_s}
\end{center}
\end{figure}

It is interesting to further explore the behavior of the model as a function of the density $\rho$ of agents in the city, which is expected to have a significant impact on the equilibrium rent, since the density plays a key role in the demand. The evolution of the equilibrium rent distribution with $\rho$ is shown on Fig.~\ref{Prix_rho_s}.
More precisely, the right panel of Fig.~\ref{Prix_rho_s} presents the rent distributions, while the left panel shows the distribution of the logarithm of the rent.
The rent distributions are well fitted by lognormal laws (right panel of Fig.~\ref{Prix_rho_s}), which simply appear as Gaussian laws when plotting the distribution of the logarithm of the rent.
House rents have been often modelled as lognormal distributions in the housing economics literature following the founding work of \cite{Case1987}, for example in mortgage default models \cite{Deng2000}. Our model provides a parcimonious framework leading to this feature.

Two main observations can be drawn from this study of the influence of density. The first one is the fact that the rent of housing increases when the density $\rho$ is increased.
This stylized fact is well-known in real cities \cite{anas98}, and is also reproduced by the standard urban economics model \cite{Fujita89}. In the simple model presented here, this phenomenon can be interpreted as the fact that a higher density $\rho$ yields a higher competition on the urban housing market, and thus higher rents.

The second observation which can be made on Fig.~\ref{Prix_rho_s} is the decreasing width
of the distribution of the logarithm of the rent, when the density $\rho$ is increased.
This can be linked to the multiplicative dynamics of the rent,
according to the following interpretation.
Let us suppose that rent $p$ is such that $p<p_s$ and $p<p_l$. For high rents,
a relative variation of the rent $p$ (multiplication by $f_r>1$ or $f_l<1$) generates a higher absolute variation of the probability $\pi_s=p/p_s$ that a tenant tries to move out of the flat (see the right panel of Fig.~\ref{prix_occup}) and of the probability $\pi_l=p/p_l$ that the rent of an empty flat decreases. These higher variations prevent the rent from increasing ($\pi_r$) or decreasing ($\pi_s$) too much. Indeed, the reactions to rent variations are quicker for higher rents, and the rent is bound to stay within a smaller domain, in logarithmic scale.

\section{Analytical approach}
\label{analytical}

In order to better understand the numerical results obtained, we propose a simplified
version of the model, that can be solved analytically.
In this simplified model, we consider that the (decimal) logarithm $x=\log_{10} p$ of the rent $p$ of housing describes a biased random walk on a one-dimensional lattice, with lattice spacing $a$, so that $x=na$.
This variable $x$ evolves by integer jumps of $n_r a=\log_{10} f_r$ in the positive direction, or $n_la=-\log_{10} f_l$ in the negative direction, due to rent variations by the owner.
In a mean-field spirit, the probability per unit time $\tilde{q}$ for the rent to be raised should be proportional to the density $\rho$ and to $\pi_r$, so that a first estimate would be $\tilde{q} = \pi_r \rho$.
Similarly, the probability $q(x)$ for the rent to be lowered is proportional to $1-\rho$ and to
\be \label{eq-pilx}
\pi_l(x)=p(x)/p_l=10^x/p_l,
\ee
leading to $q(x) = (1-\rho) \pi_l(x)$.
Note that only the probability to decrease the rent depends on the rent, through its logarithm $x$.

\begin{figure}[t]
\begin{center}
\begin{tabular}{cc}
\includegraphics[width=5.5cm]{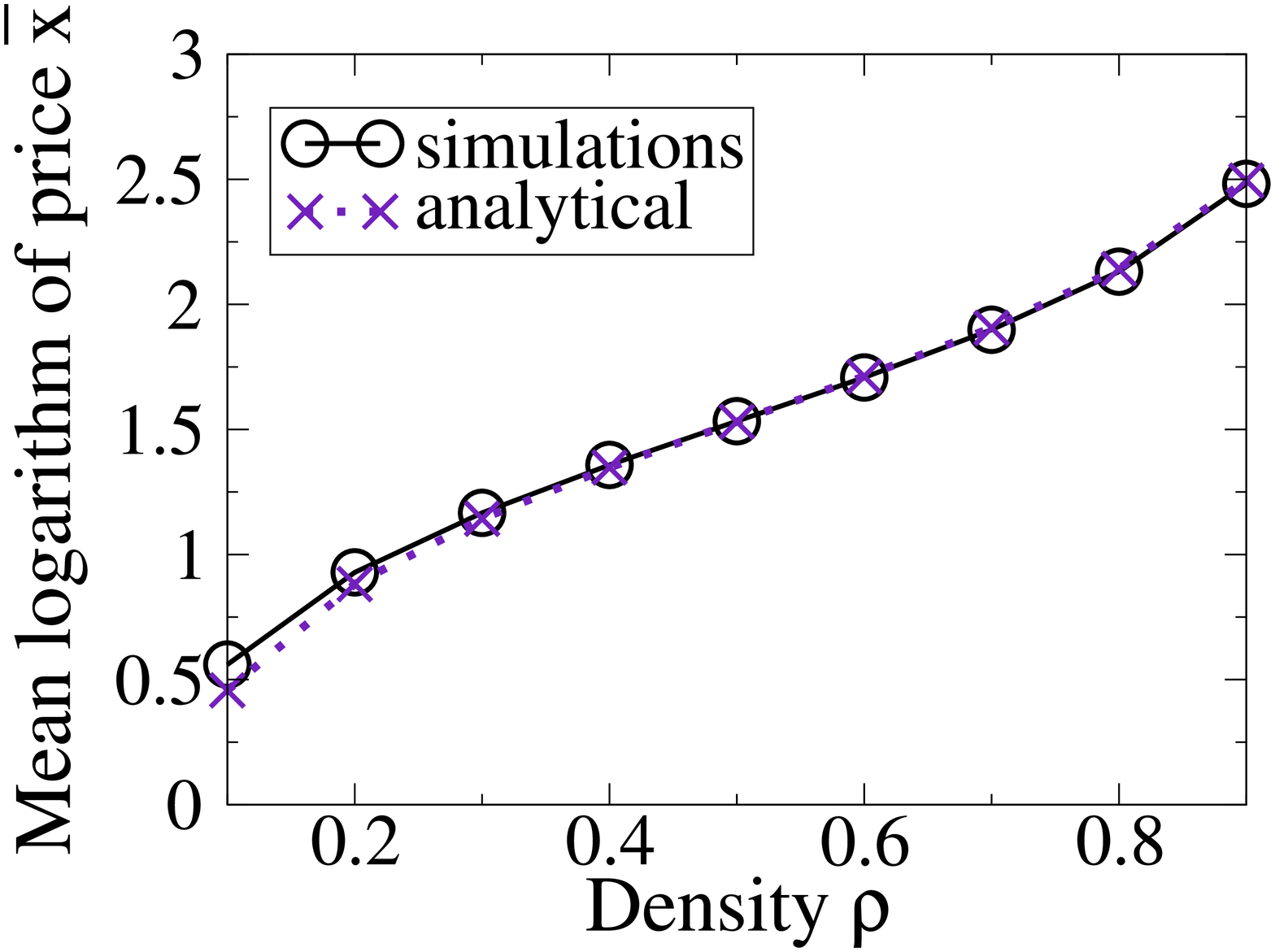} &
\includegraphics[width=5.5cm]{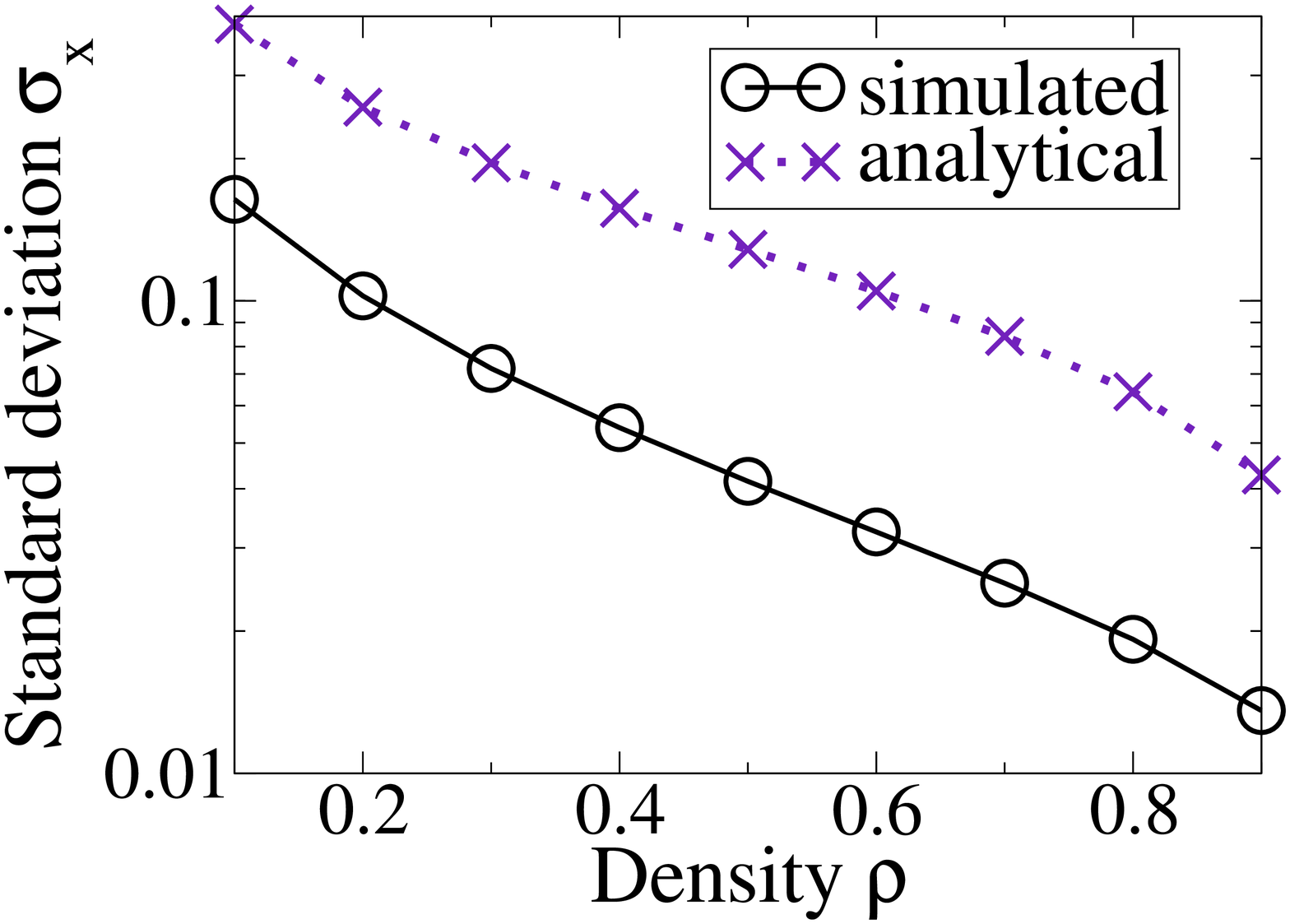} \\
\includegraphics[width=5.5cm]{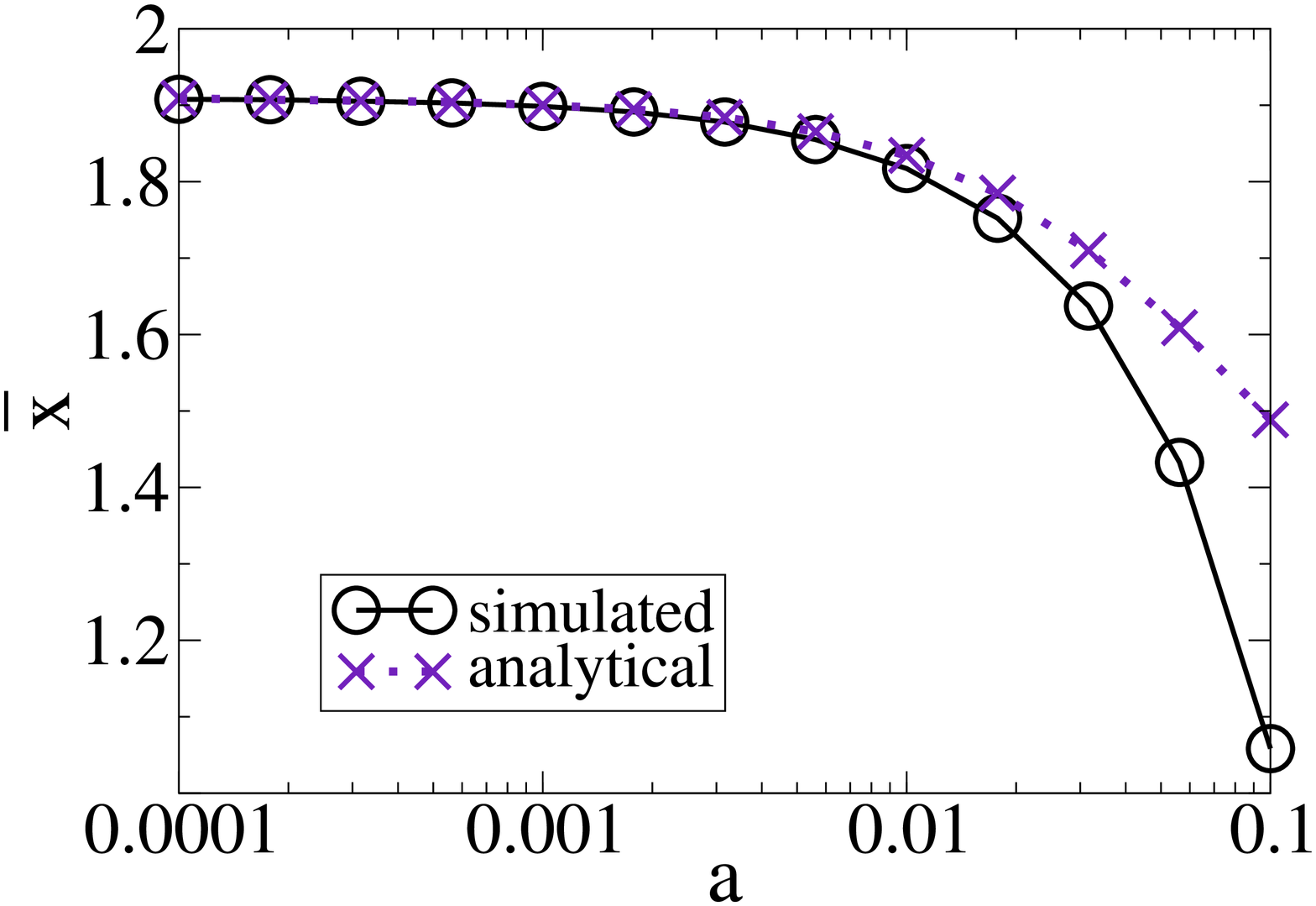} &
\includegraphics[width=5.5cm]{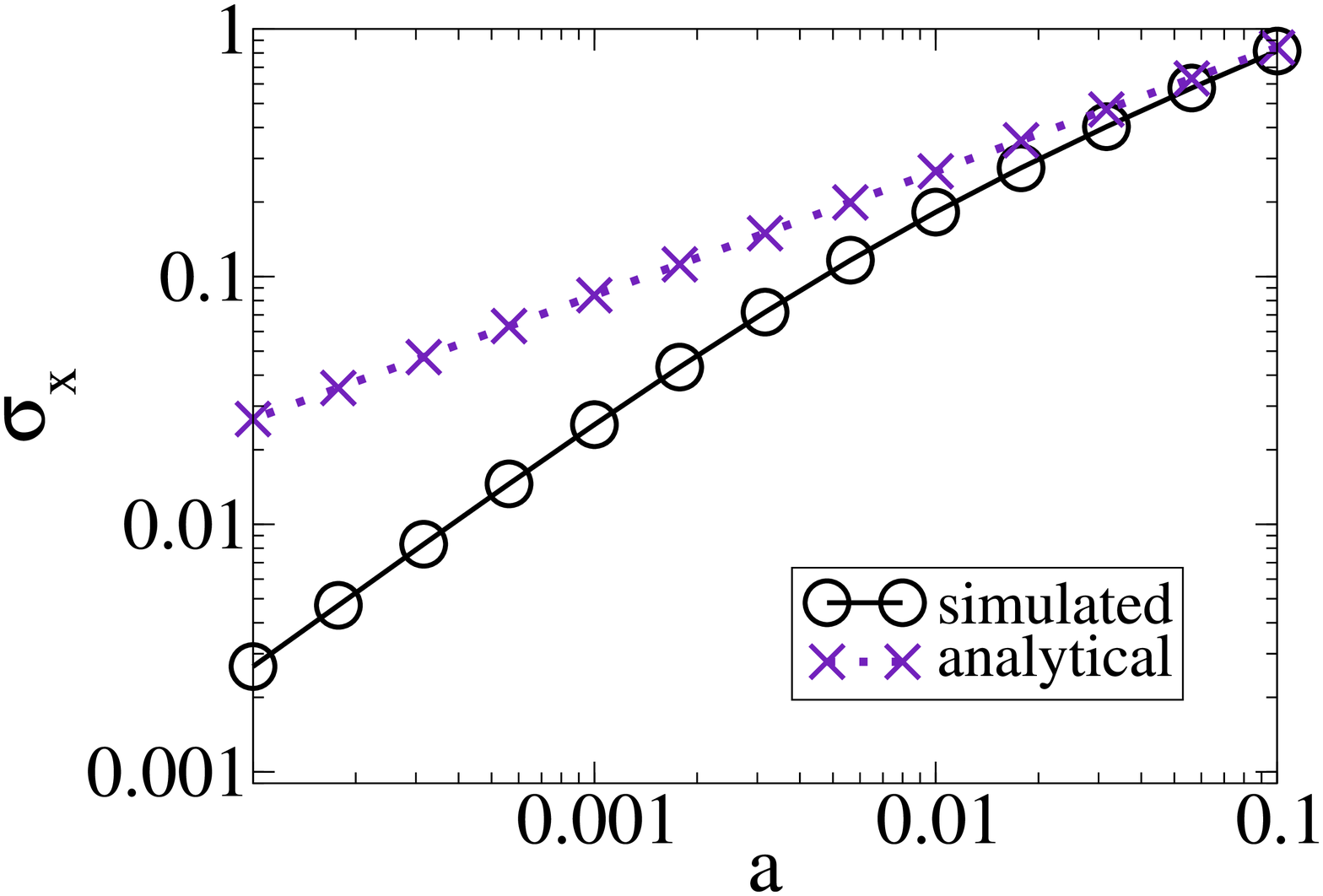}
\end{tabular}
\caption[Mean and standard deviation of rent]{Evolution of the mean $\bar{x}$ (left column) and of the standard deviation $\sigma_x$ (right column) of the decimal logarithm $x$ of rent as functions of the density $\rho$ (top row) and the lattice spacing $a$ (bottom row). Other parameters: same as in Fig.~\ref{EvoPmean}.}
\label{moysigma}
\end{center}
\end{figure}

To take into account the fact that a flat remains occupied or empty during several successive time steps, so that the steps of the random walk should be correlated, we
introduce the mean number $n_o$ of rent increases during a period when a flat is occupied, and the mean number $n_v$ of rent decreases during a period when a flat is vacant.
We then consider that the variable $x$ performs effective steps $n_o n_r a$ and $n_v n_la$
(instead of $n_r a$ and $n_la$), assumed to be statistically independent.
In order to maintain the same average rate of rent increase and decrease, we divide
the probabilities $\tilde{q}$ and $q(x)$ respectively by $n_o$ and $n_v$, yielding
\be \label{eq-rates-q}
\tilde{q} = \frac{\pi_r \rho}{n_o}, \qquad 
q(x) = \frac{\pi_l(x)(1-\rho)}{n_v}.
\ee
Equating the mean increase rate $\tilde{q}n_o n_r$ and decrease rate
$q(x_{\rm{eq}})n_v n_l$ of $x$, the equilibrium rent satisfies
\be
q(x_{\rm{eq}}) = \tilde{q}\,\frac{n_r n_o}{n_l n_v}.
\ee
Using Eqs.~(\ref{eq-pilx}) and (\ref{eq-rates-q}),
we find the equilibrium rent $p_{\rm{eq}}=10^{x_{\rm{eq}}}$:
\be \label{peq}
p_{\rm{eq}} = \frac{n_r}{n_l}\; \frac{\pi_r p_l \rho}{ 1-\rho}.
\ee
Note that the equilibrium rent is independent of the phenomenological parameters
$n_o$ and $n_v$.

We now try to estimate the parameters $n_o$ and $n_v$.
When a flat is occupied, at each step there is on average a probability approximately equal to $\frac{1}{2}\pi_s=p_{\rm{eq}}/(2p_s)$ that the tenant moves to another flat, assuming that all rents are close to the equilibrium rent $p_{\rm{eq}}$. Thus the probability that a tenant has not changed flats after $n$ steps is
\be
F_n = \left( 1-\frac{p_{\rm{eq}}}{2 p_s}\right)^n,
\ee
and the probability that a tenant leaves the flat after exactly $n$ steps is
$f_n=F_{n-1}-F_n$, yielding
\be \label{eq-fn}
f_n = \frac{p_{\rm{eq}}}{2 p_s} \left( 1-\frac{p_{\rm{eq}}}{2 p_s} \right)^{n-1}.
\ee
Hence, $n_o$ can be estimated as the average number of steps a tenant
stays in a given flat, times the probability $\pi_r$
that the rent is raised at each time step:
$n_o=\pi_r\sum_n nf_n$. 
Using Eqs.~(\ref{peq}) and (\ref{eq-fn}), this estimate leads to
\be
n_o = \frac{2 p_s n_l (1-\rho)}{p_l n_r\rho}.
\ee
The computation of $n_v$ follows the same scheme. The probability for a tenant to change flats is $p_{\rm{eq}}/(2p_s)$ at each time step, so that the probability for a vacant flat to welcome a tenant is
\be
\pi_v = \frac{p_{\rm{eq}}\, \rho}{2 p_s (1-\rho)}
\ee
(we assume in this simplified argument that the density is not too high, so that
$\pi_v$ remains less than $1$).
The probability that the flat remains vacant is $1-\pi_v$ at each time step, and $(1-\pi_v)^n$ for $n$ time steps. Following the same lines as for $n_o$, we obtain
$n_v=\pi_l(x_{\rm{eq}})/\pi_v$ with $\pi_l(x_{\rm{eq}})=p_{\rm{eq}}/p_l$, eventually yielding
\be
n_v = \frac{2 p_s (1-\rho)}{p_l \rho}.
\ee
Note that the parameters $n_o$ and $n_v$ depend on $p_s$, which is consistent with the results presented in the right panel of Fig.~\ref{prix_occup}.
Indeed, we find both in the numerical simulations and in the analytical model
that $p_s$ has an influence on rent fluctuations, but not on the mean rent
(which in the analytical model is independent of $n_o$ and $n_v$).

To determine the distribution of $x$, we first write the master equation
governing the probability $P_n(t)$ that the random walk is at a position
$n=x/a$ at time $t$, using the probability rates $\tilde{q}$ and $q(x)$
given in Eq.~(\ref{eq-rates-q}):
%\bea
%\frac{dP_n(t)}{dt} &=& -(\tilde{q} + q_n) P_n(t)
%+ \tilde{q} P_{n-n_o n_r}(t)\nonumber \\
%&& + q_{n+n_v n_l} P_{n+n_v n_l}(t),
%\eea
\be
\frac{dP_n(t)}{dt} = -(\tilde{q} + q_n) P_n(t) + \tilde{q} P_{n-n_o n_r}(t) + q_{n+n_v n_l} P_{n+n_v n_l}(t),
\ee
with the notation $q_n=q(na)$.
If $a$ is small, the distribution essentially appears as continuous, and $P_n(t)$
can be described by a continuous distribution $\varphi(x,t)$ through the relation
$P_n(t)=a\varphi(na,t)$.
Expanding the different terms to second order in $a$, we obtain the Fokker-Planck equation
\be
\frac{\partial \varphi(x,t)}{\partial t} = -\frac{\partial  J(x,t)}{\partial x}
\ee
with 
\bea
J(x,t) &=& a[ n_o n_r \tilde{q} - n_v n_l q(x)] \varphi(x,t)\nonumber \\
&& - \frac{a^2}{2}\frac{\partial}{\partial x}
\Big[ \big( n_o^2 n_r^2 \tilde{q} + n_v^2 n_l^2 q(x) \big)\varphi(x,t) \Big].
\eea
Searching for a stationary solution with zero flux, we solve the equation $J(x,t)=0$ \cite{VanKampen} which yields, to leading order in $a$, a stationary distribution of the form
\be
\varphi(x) \propto e^{\frac{1}{a}\, g(x)},
\ee
where $g(x)$ is a function that can be determined explicitly, and that exhibits
a maximum for $x_{\rm{eq}}$.
Given that $a$ is small, $\varphi(x)$ is dominated by the vicinity of the maximum of $g(x)$
and one can expand $g(x)$ around $x_{\rm{eq}}$, to second order in $x-x_{\rm{eq}}$,
leading to
\be
\varphi(x) = \frac{1}{\sigma\sqrt{2\pi}} \exp \left[\frac{(x-x_{\rm{eq}})^2}{2\sigma^2} \right]
\ee
with
\be
\sigma = \sqrt{\frac{a}{2\ln 10}(n_o n_r + n_v n_l)}.
\ee
Coming back to the rent variable $p$,
we thus recover a lognormal law,
similarly to the results obtained from numerical simulations.
Fig.~\ref{moysigma} shows that this simple analytical model is in good agreement with the numerical simulations as far as the equilibrium rent is concerned, which is a non-trivial result.
The analytically predicted standard deviation presents a qualitatively correct
behavior as a function of density (namely, it decreases when density is increased),
but its magnitude is overestimated by a factor of the order of $3$.
This discrepancy may be partly explained by the fact that the random walk
performed by the logarithm of the rent in the numerical simulations
is anticorrelated, meaning that a jump of characteristic length $n_on_r$ to the right 
is followed by a jump of characteristic length $n_vn_l$ to the left, and reciprocally.
This is not the case with the random walk we consider here, in which the effective jumps
of length $n_o$ and $n_v$ are uncorrelated, allowing several consecutive jumps
in the same direction. This decorrelation is expected to make the rent dispersion
larger than in the simulated model.

{\color{black} An interesting test of this model would be to fit empirical data of rent distributions looking like lognormal distributions. Focusing for simplicity on our analytical model, measurement of $x_{\rm{eq}}$ and $\sigma$ on such empirical data would determine respectively $\pi_r p_l n_r / n_l$ and $a p_s n_l / p_l$, supposing that the vacancy rate fixes the density $\rho$. Some additional knowledge of the rental housing market and its dynamics is then needed to find acceptable values of the characteristic rents $p_l$ and $p_s$, and of the parameters describing the "microscopic" dynamics of rent: $n_r$, $n_l$, $\pi_r$ and $a$. Such microscopic parameters may however be difficult to determine.}

\section{Conclusion and perspectives}

We have studied a simple model of urban housing market, focusing on the distribution
of rents (or rents).
The mechanisms of rent formation are simple and intuitive: flats which are highly demanded have an increasing rent, whereas flats for which agents have a low interest have a decreasing rent.
The balance between supply and demand resulting in the equilibrium rent distribution can be seen as a bargaining between tenants and landlords, where very few hypotheses are introduced concerning the respective bargaining powers of both categories of agents.
From a physics viewpoint, one can see that non-linearities encoded in behavioral rules
(through the fact that the probability to change flat for tenants, or to lower the rent
for landlords, depends on the rent) play a major role in allowing for a saturation of rent dispersion.
Without these non-linearities, the rent distribution would continuously spread,
just like in the case of an unbounded random walk.

The main specificity of this work when compared to economic literature is the simple stochastic behavior of agents, whereas most models in economics use rules which are at least partly deterministic.
As a result of the use of stochastic behavior rules for agents, the equilibrium of our model is different from the equilibria commonly encountered in the economic litterature, which are usually static in some sense: once rents have reached an equilibrium distribution, firms do not have an incentive to change them, so that they do not evolve anymore \cite{Burdett1983, Wheaton1990}. In our model, the equilibrium distribution of rents is stationary even though rents still evolve at equilibrium. However, the system has reached a stationary state, where moves of tenants and rent adjusments by landlords are such that they do not change the rent distribution, a result which is reproduced by our analytical approach.

This work also emphasizes the importance of the density of agents, which plays a key role in the determination of both the equilibrium rent and the amplitude of fluctuations.
The equilibrium rent increases with density, while the relative amplitude of fluctuations
(given by the standard deviation of the logarithm of the rent) decreases with density.
This tends to validate an hypothesis made in an earlier work \cite{epl2011},
of a direct positive relationship between equilibrium rent and density.
Numerical results are qualitatively well reproduced by a model of biased random walk
with uncorrelated steps, providing a simple understanding of the dynamics of the simulated model. A quantitative agreement is even found for the equilibrium rent.

Extensions of this work could be considered in several directions.
Our model could be considered as a labor market model with on-the-job search (see \cite{Burdett1998}). There is however no equivalent of frictional unemployment: the vacancy rate is non zero, but all agents have a home, which corresponds to all workers being employed on the labor market.
A further research direction could be to study the labor market by modifying our model to allow for unemployment (that is, agents leaving their jobs even before having found another one). Such a work could attempt to explain why wage dispersion is larger in real data than predicted by search models \cite{Hornstein2011}.
More immediately, introducing new
ingredients in the model would make it slightly more realistic.
For instance, one could introduce a cost of move $\Delta p$: agents would move
into a new flat only if $p_e-p>\Delta p$.
This cost of move should have a friction-like effect on the housing market, and is expected to increase the width of the distribution of rents.
In addition, differentiation between flats could be introduced, which might allow
for a more precise comparison with the literature on urban economics, where flats
are differentiated for instance by their distance to the city center \cite{Fujita89}.

On the statistical physics side, it would be interesting to investigate in more details the connection with ratchet models and dichotomous Markov noise \cite{bena06}.
Indeed, the dynamics of the rent of a given flat switches between
exponential increase and decrease phases, depending on whether the flat is occupied or empty.
This situation is similar in spirit to ratchet models with dichotomous Markov noise,
in which the dynamics of a particle randomly switches between two different regimes,
corresponding for instance to two different potential energy profiles \cite{bena06}.
There are however some subtleties in our model which make it different from a standard ratchet
model.
First of all, the change of state (occupied or empty) of a given flat is not strictly speaking
a random noise, but results from the $N$-body dynamics of the whole system.
However, this first difficulty could be circumvented by considering a self-consistent noise
in a mean-field spirit. More importantly, the probability that an agent changes flat depends
on the corresponding rent, which introduces a non-trivial coupling between rents
and occupancy. Hence standard methods from ratchet theory cannot be straightforwardly
applied, and extensions would be necessary.

\section*{References}
\bibliographystyle{unsrt}
\bibliography{biblio}

\end{document}